\documentclass[twocolumn,nofootinbib,amsmath,prl,aps,superscriptaddress,tightenlines,preprintnumbers]{revtex4}

\usepackage{graphicx}
\usepackage{braket}
\usepackage[usenames]{color}
\usepackage{latexsym}
\usepackage{bm}
\usepackage{mathrsfs}
\usepackage[caption=false]{subfig}

\usepackage{tikz}
\usetikzlibrary{arrows,decorations.pathmorphing,backgrounds,positioning,fit,petri,automata,shadows,calendar,mindmap,
decorations.markings,calc}

\usepackage{braket,slashed,bm,bbm}
\usepackage[utf8]{inputenc}
\usepackage{graphicx}
\usepackage{feynmp-auto}

\def\bea{\begin{eqnarray}}
\def\eea{\end{eqnarray}}

\begin{document}

\newcount\hour \newcount\minute
\hour=\time \divide \hour by 60
\minute=\time
\count99=\hour \multiply \count99 by -60 \advance \minute by \count99
\newcommand{\mydate}{\ \today \ - \number\hour :00}

\title{\! \! The Neutrino Option}

\author{Ilaria Brivio and Michael Trott,\\
Niels Bohr International Academy and Discovery Centre,\\
 Niels Bohr Institute, University of Copenhagen,
Blegdamsvej 17, DK-2100 Copenhagen, Denmark
}

\begin{abstract}
The minimal seesaw scenario can radiatively generate the Higgs potential to induce electroweak
symmetry breaking while supplying an origin of the Higgs vacuum expectation value from an underlying Majorana scale.
If the Higgs potential and (derived) electroweak scale have this origin, the heavy $\rm SU(3) \times SU(2) \times U(1)_Y$ singlet states are expected to reside
at $m_N \sim 10-500 \, {\rm PeV} $ for couplings $|\omega| \sim 10^{-4.5}-10^{-6}$ between the Majorana sector and the Standard Model. In this framework, the usual challenge of the
electroweak scale hierarchy problem with a classically assumed potential
is absent as the electroweak scale is not a fundamental scale. The new challenge  is the need to generate or accommodate PeV Majorana mass scales
while simultaneously suppressing tree-level contributions to the potential in ultraviolet models.
\end{abstract}

\maketitle
\newpage

\paragraph{\bf I. Introduction.}
The Standard Model (SM) provides a successful description of many
particle physics measurements but
does not explain the experimental evidence for
dark matter and neutrino masses.
These facts argue for extensions of the SM that experience
some form of decoupling in an Effective Field Theory (EFT) setting \cite{Appelquist:1974tg}
so that the corrections to the SM in an effective field theory (SMEFT) framework
are small perturbations.

Minimal (viable) extensions of the SM are usually beset with an inability to address the Electroweak (EW) scale hierarchy problem.
A SMEFT statement of which is that threshold corrections to $H^\dagger H$ can be generated by
integrating out sectors extending the SM proportional to large scales $\Lambda \gg m_h$.
Without parameter tuning the Higgs mass is expected to be proximate to the cut off scale of the theory for this reason.
Symmetries, such as supersymmetry, can suppress these threshold corrections.
This is frequently done while assuming the EW scale is a fundamental parameter and the SM Higgs potential has
a classical form that leads to $\rm SU(2)_L \times U(1)_Y \rightarrow U(1)_{em}$. As a result,
one generally expects new particles related to stabilizing symmetries, such as superpartners, to be around the TeV scale and
found at LHC. Unfortunately, this has not (as yet) occurred.

In this paper we do not adopt the assumptions that the EW scale and Higgs potential are fixed classically without a dynamical origin.
We develop an alternative approach using the minimal seesaw scenario \cite{Minkowski:1977sc,GellMann:1980vs,Mohapatra:1979ia,Yanagida:1980xy} as an extension of the SM.
The idea is that the Higgs mass and potential are generated by threshold corrections from
the Majorana \cite{Majorana:1937vz} sector and the anomalous breaking of scale invariance in the Coleman-Weinberg (CW) potential \cite{Coleman:1973jx}.
The result is an origin of the observed neutrino masses within a SM extension
that avoids parameter tuning. An origin for the EW scale is introduced
that suggests a new perspective on the EW scale hierarchy problem, modifying the
usual concerns that lead to expectations of TeV scale new states into an alternate framework.
The purpose of this paper is to demonstrate this possibility --
``the neutrino option''.
\\
The scenario presented is falsifiable and depends in a sensitive manner on the measured value of the
top quark and Higgs masses, and the detailed neutrino mass spectrum.
Higher order Renormalization Group Equations (RGEs) and threshold matching calculations are also
critical in this scenario. This approach strongly motivates more theoretical and experimental progress
in all of these areas in order to falsify or confirm this possible origin of the Higgs potential and EW scale.

\paragraph{\bf II. Seesaw model and threshold corrections.}
We use the seesaw formalism of Refs.~\cite{Broncano:2002rw,Elgaard-Clausen:2017xkq}.
The extension of the SM Lagrangian with $p,q=\{1,2,3\}$ singlet fields $N_p$ is given by
\bea\label{basicL}
2 \,\mathcal{L}_{N_p}&=&  \overline{N_p} (i\slashed{\partial} - m_{p})N_p - \overline{\ell_{L}^\beta} \tilde{H} \omega^{p,\dagger}_\beta  N_p,  \\
&-&  \overline{\ell_{L}^{c \beta}} \tilde{H}^* \, \omega^{p,T}_\beta N_p - \overline{N_p} \, \omega^{p,*}_\beta \tilde{H}^T \ell_{L}^{c \beta}  - \overline{N_p}\, \omega^p_\beta \tilde{H}^\dagger \ell_{L}^\beta.\nonumber
 \label{LNa}
\eea
The $\omega^p_\beta = \{x_\beta,y_\beta,z_\beta\}$ are each complex vectors in flavour space.
These vectors have absorbed the Majorana phases $\theta_p$.
The mass eigenstate Majorana fields are defined such that they satisfy the Majorana condition \cite{Majorana:1937vz}: $N_p^c= N_p$.
These fields are related to chiral right handed fields $N_R$ that are singlets under $\rm SU(3) \times SU(2)_L \times U(1)_Y$ as \cite{Broncano:2002rw}
\bea
N_p = e^{i \theta_p/2} \, N_{R,p} + e^{- i \theta_p/2} \, (N_{R,p})^{c}.
\eea
The superscript $c$ stands for charge conjugation; defined on a four component Dirac
spinor as $\psi^c  = - i \gamma_2 \, \gamma_0 \overline{\psi}^T$. Integrating out
the seesaw model at tree level, the results can be mapped to the SMEFT up to dimension seven \cite{Elgaard-Clausen:2017xkq}. The threshold corrections of interest
lead to dimension two and four terms.
They come about due to integrating out the heavy $N_p$ states at one loop
and running down to the scales that are used to experimentally probe the SM states and interactions in the EW vacuum.
\begin{figure}
\includegraphics[width=0.45\textwidth]{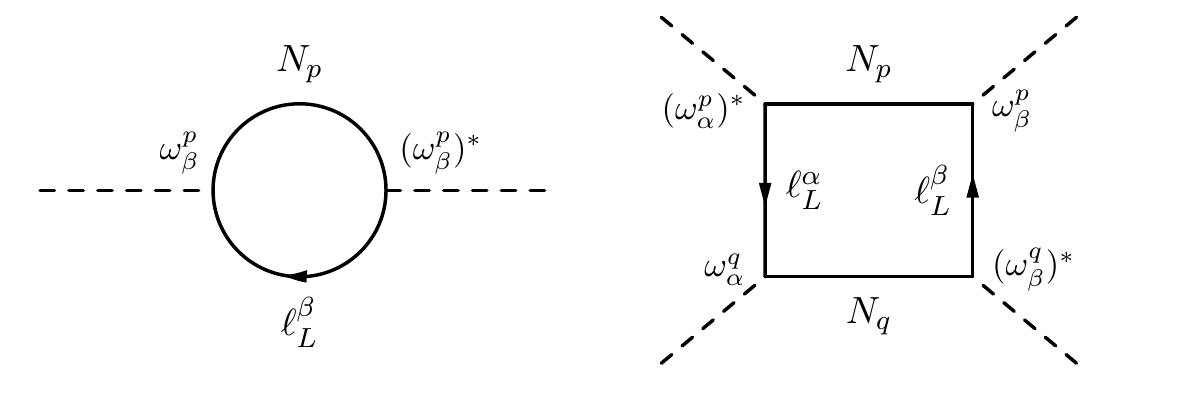}
\caption{\label{fig:threshold} One loop corrections matching onto and generating the Higgs potential
at the scale(s) $m_p$ in the seesaw model.}
\end{figure}
The diagrams of Fig.\ref{fig:threshold} give a threshold matching to the Higgs potential terms
\bea\label{potentialterms}
V(H^\dagger \, H) = -\frac{m_p^2 |\omega_p|^2}{16 \, \pi^2}\, F_1 \, H^\dagger H
-\lambda_{pq} \, F_2 \, (H^\dagger H)^2,
\eea
where
\bea
F_1 &=& 1 + \log \frac{\mu^2}{m_p^2}, \\
F_2 &=& 1 - \frac{m_p m_q \, \log \frac{m_p^2}{m_q^2} + m_q^2 \log \frac{\mu^2}{m_q^2}-m_p^2
\log \frac{\mu^2}{m_p^2}}{(m_p^2-m_q^2)},\label{basicmatching}
\eea
and $\lambda_{pq} = 5 \, (\omega_q \cdot \omega^{p,\star})(\omega_p \cdot \omega^{q,\star})/(64 \,\pi^2)$.
Here the repeated indicies are summed over and the results can be compared to
past results in Ref.~\cite{Casas:1999cd,Bambhaniya:2016rbb}.
We have used dimensional regularization in $d = 4 - 2 \epsilon$ dimensions and
$\rm \overline{MS}$ here and below. The counterterms of the SM and the full theory including the $N_p$ states
cancel the $\epsilon$ divergences in each case. The mismatch of the SMEFT Lagrangian and the full theory in the limit
$p^2/m_p^2 \rightarrow 0$
(with kinematic invariants denoted $p^2$) defines the threshold matching. The renormalization scale dependent
logs in the result can be neglected when also neglecting the (net) two loop effects due to running between the threshold matchings.

Considering the parameterization of the Higgs potential ($V_c$) as
\bea\label{VCeqn}
V_c(H^\dagger H) = - \frac{m^2}{2} \,(H^\dagger \, H) + \lambda \, (H^\dagger \, H)^2,
\eea
neglecting the effect of running down from the scale(s) $\mu = m_p$, and
assuming the mass differences are parametrically smaller than intrinsic mass scales
in the Majorana sector ($(m_q^2-m_p^2)/m_{p,q}^2 < 1$) we find
\bea\label{treelevel}
\Delta \, m^2 = m_p^2 \frac{|\omega_p|^2}{8 \, \pi^2}, \quad \Delta \, \lambda =-5 \frac{(\omega_q \cdot \omega^{p,\star})(\omega_p \cdot \omega^{q,\star})}{64 \,\pi^2}.
\eea
Note that the $\lambda$ threshold corrections can be subdominant to other
quantum corrections in the full CW potential
in the parameter space of interest where $|\omega_p| \ll 1$ and $m_p \gg 246 \, {\rm GeV}$.
\paragraph{\bf III. Induced CW potential.}
The threshold corrections to $H^\dagger H$
can be naturally dominant in defining the Higgs potential below the scales
$\mu \simeq m_p$. The reason is that the SM is classically scale invariant in the limit
that the vacuum expectation value (vev) of the Higgs $v \rightarrow 0$ \cite{Coleman:1973jx,Weinberg:1978ym,tHooft:1979rat,Bardeen:1995kv}
 ($m \rightarrow 0$
in Eqn.~\ref{VCeqn}).
This point of enhanced symmetry is anomalous,
even before its soft breaking by the threshold matching.
However, the additional SM breakings of scale invariance through
quantum corrections are associated with dimensionful parameters that are smaller than $m_p^2$
in a consistent version of this scenario at the threshold matching scale.

A breaking of the scaleless limit of the SM is due to
QCD, which generates the scale $\Lambda_{QCD}$
by dimensional transmutation \cite{Coleman:1973jx} at low scales as $(\Lambda_{QCD}/\mu)^{b_0} = {\rm Exp} \left[-8 \, \pi/\hbar g_3^2(\mu)\right]$
where $b_0 = 11 - (2/3)\, n_f$ \cite{Politzer:1973fx,Gross:1973id}.
   The quark masses
that result lead to $V_c$ contributions such as
\bea
\Delta m^2 = \frac{N_c \, y_t^2 \, \Lambda_{QCD}^2}{32 \, \pi^2}\left(1+ 3 \, \log \left[\frac{\mu^2}{\Lambda_{QCD}^2}\right] \right) + \cdots
\eea
which subsequently induces a vev for the Higgs, leading to gauge boson masses $\propto \Lambda_{QCD}$.
As we are assuming $m_p^2 |\omega_p|^2 \gg \Lambda_{QCD}^2$ (for each p) these contributions are naturally subdominant
for $H^\dagger H$, and anyway, at the threshold matching scale we consider, the QCD coupling
has run to scales such that $g_3(m_p) < 1$.
Renormalization of the CW potential
also introduces an anomalous breaking of scale invariance. Consider
defining $V_{CW}(\langle H^\dagger H \rangle)$ as the one loop CW potential
expanding around the scaleless limit of the SM while neglecting the threshold corrections.
The standard result \cite{Coleman:1973jx,Weinberg:1978ym,tHooft:1979rat}
can be minimized via
$\partial V_{CW}/\partial \langle H^\dagger H\rangle  =0$.
The vev scale obtained is exponentially separated from the
renormalization scale. This scale is associated with the
asymptotic nature of the perturbative expansions used in constructing the
CW potential, that also predict $S$-matrix
elements that are used to fix SMEFT Lagrangian parameters.
This scale can be either suppressed or enhanced depending on
the net sign of the quantum correction in the CW potential,
and a suppression is consistent with an EFT analysis.

In summary, the soft breaking of the scaleless limit of the SM\footnote{Previous studies of the CW potential in this scaleless limit (not advocating
the Neutrino Option) include Refs.~\cite{Bardeen:1995kv,Hempfling:1996ht,Meissner:2006zh,Foot:2007iy,Shaposhnikov:2008xi,Lindner:2014oea,
Helmboldt:2016mpi,Clarke:2015gwa}.
We have introduced
a hard breaking of scale invariance due to $m_p$ in Eqn.~\ref{basicL}. If such masses were spontaneously generated the breaking of scale invariance
would be completely spontaneous and introduce a dilaton \cite{Gildener:1976ih} in the spectrum. This scenario is beyond the scope of this work. We thank D. McGady for conversations on this point.}
is such that the
threshold corrections to $H^\dagger H$ due to integrating out the $N_p$ states
can be a dominant contribution to $V_{CW}$ fixing a high scale boundary condition for the Higgs potential. This occurs for interesting parameter
space when tuning of the threshold corrections against bare parameters is
avoided expanding around the classically scaleless limit of the SM Lagrangian.

\paragraph{\bf IV. Running down to the scale $\mu = \hat{m}_t$.}
We assume that the Higgs potential is (dominantly) given by Eqn.~\ref{treelevel}
when integrating out the Majorana sector. This condition can be obtained requiring (i) smaller breaking of scale invariance in the Higgs sector
and (ii) the bare tree-level Higgs potential at the scale $m_p$ is negligible compared to the threshold contributions.
The realization of these conditions in a full UV model represents a challenge alternative to the usual hierarchy problem.
In such a scenario,
a non-trivial consistency condition is the successful generation
of the SM potential at lower energy scales and field values.
The measured masses
of the SM states and their couplings ensures this can occur in a very non-trivial fashion.
Note also that despite the fact that the seesaw boundary condition fixes $\lambda <0$,
this coupling can still run to positive values at lower scales as it is not multiplicatively renormalized.

To demonstrate that a successful lower scale phenomenology can result
from these seesaw boundary conditions, we take $V_c(H^\dagger \, H)$ to be fixed to
$m^2(m_p) \equiv \Delta m^2$ and $ \lambda(m_p) \equiv \Delta \lambda$.
The parameters $m^2$ and $\lambda$ are then run down according to the coupled SM
RGEs.
The $\beta$-functions for running above the top mass are introduced as $\beta (x) = (4 \pi)^2 \, d x/d \ln\mu^2$
and are taken (at leading order) from the summary in Ref.~\cite{Buttazzo:2013uya} as
\bea
\beta(g_Y^2) &=&  g_Y^4 \frac{41}{6},
 \quad \! \! \! \!
\beta(g_2^2) =  g_2^4 \left(-\frac{19}{6}\right), \quad \! \! \! \!
\beta(g_3^2) =  g_3^4 (-7), \nonumber \\
\beta(\lambda) &=& \left[\lambda\left(12 \lambda + 6 Y_t^2-\frac{9}{10}(5g_2^2+\frac{5 \,g_Y^2}{3})\right) \right. \\
&-& \left. 3Y_t^4+\frac{9}{16}g_2^4+\frac{3}{16}g_Y^4+\frac{3}{8}g_Y^2 g_2^2\right],  \nonumber  \\
\beta(m^2) &=&  m^2 \left[6\lambda+3Y_t^2-\frac{9}{20}(5g_2^2+\frac{5 \,g_Y^2}{3})\right]\,, \\
\beta(Y_t^2) &=& Y_t^2 \left[\frac{9}{2}Y_t^2-8g_3^2-\frac{9}{4}g_2^2-\frac{17}{12}g_Y^2\right].
\eea
Here $g_2$, $g_3$ are the coupling constants of the $\rm SU(2)_L$ and
$\rm SU(3)_c$ gauge groups, while $g_Y$ the coupling of the $\rm U(1)_Y$ group.
$Y_i = \sqrt{2} \, m_i/v$ is the Yukawa coupling of a fermion to the Higgs
with $v^2 =1/\sqrt{2} \hat{G}_F$.
Contributions proportional to $Y_b$ and $Y_\tau$ have been neglected.
The differential system is solved by fixing the boundary conditions to be
\begin{align}
 \lambda(m_p) &= -n^2\frac{5}{64\pi^2}|\omega|^4, \quad
 m^2(m_p) = \frac{n |\omega|^2}{8\pi^2} \, m_p^2,\\
 \hat{Y}_t(m_t) &= Y_t^{(0)}+ Y_t^{(1)}(m_t) = 0.9460\,, \\
 \hat{g}_Y(m_t) &= g_Y^{(0)}+ g_Y^{(1)}(m_t)= 0.3668\,,\\
 \hat{g}_2(m_t) &= g_2^{(0)}+ g_2^{(1)}(m_t) = 0.6390\,,\\
 \hat{g}_3(m_t) &= 1.1671 \,,
\end{align}
for different choices of $m_p$ and $|\omega|$ which
approximate to one common universal scale and coupling in what follows.
\begin{figure}
 \includegraphics[width=0.5\textwidth]{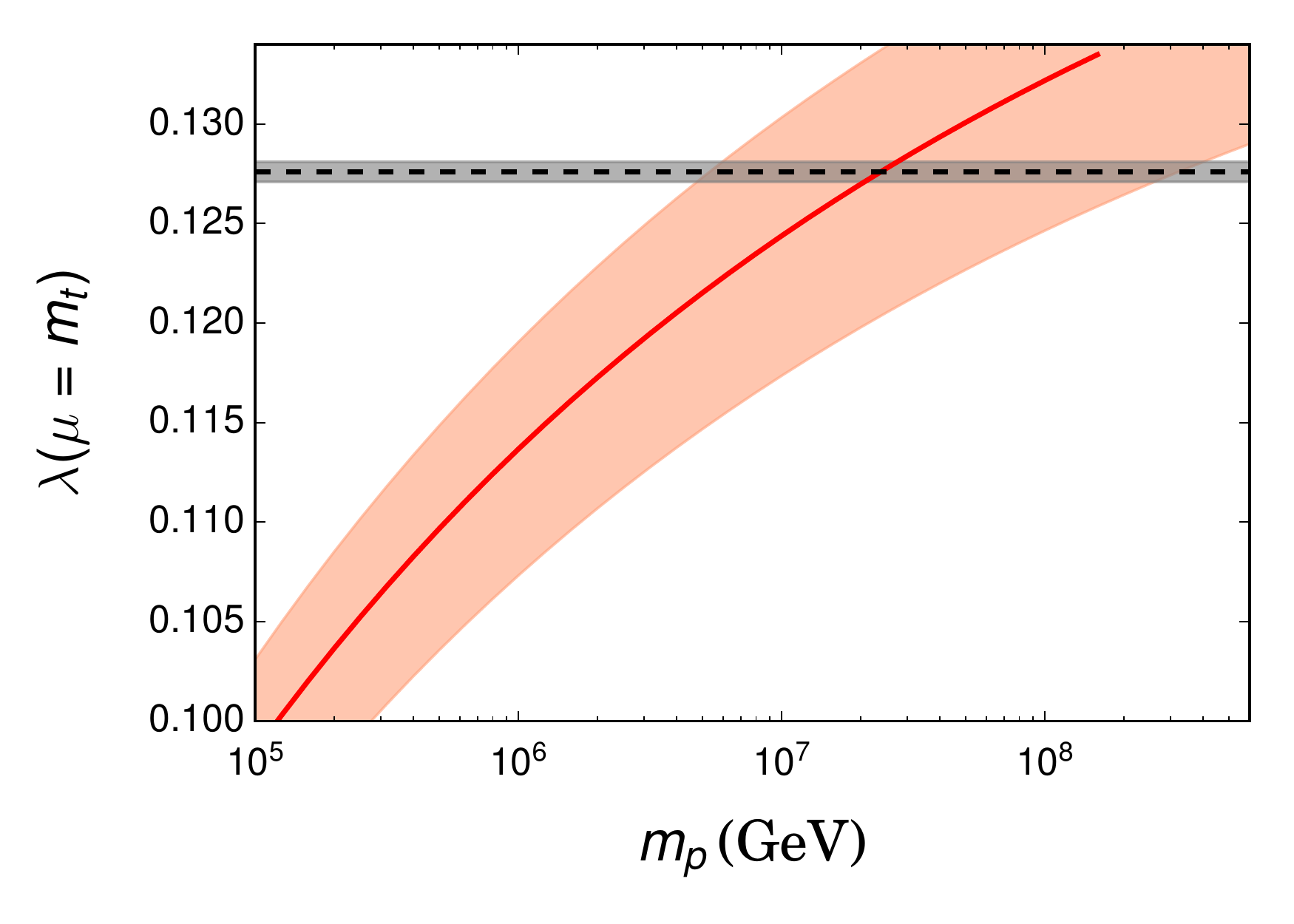}
 \caption{\label{fig:lambda_vs_M} Value of the parameter $\lambda$ extrapolated at the scale $\mu=\hat{m}_t$ as a
 function of the heavy neutrino mass scale $m_p$. The red line assumes $\hat{m}_t=173.2$~GeV. The surrounding band
 corresponds to varying $\hat{m}_t$ between 171 and 175~GeV (a $2 \sigma$ error variation \cite{Olive:2016xmw}).
 The dashed line and surrounding band indicates the value of $\lambda(m_t)$ consistent with the measured Higgs mass
 and its percentage error \cite{Aad:2015zhl}.}
\end{figure}
The number of heavy neutrino species has been denoted with $n$ and fixed to $n=3$.
The $x^{(0)}$ and $x^{(1)}(\mu)$ stand for the tree and one-loop level
contribution in the SM and we have denoted as hatted quantities observables inferred from measured input parameters.
The analytic expressions for the tree-level definitions used are standard
for the $\{\hat{m}_{W}, \hat{m}_Z,\hat{G}_F\}$ input parameter set.
The central values of the numerical inputs used are
\bea
\{\hat{m}_Z,\hat{m}_W,\hat{m}_t,\hat{m}_h\}&=&\{91.1875,80.387,173.2,125.09\}, \nonumber
\eea
in GeV units, $\hat{G}_F =1.1663787 10^{-5} \, {{\rm GeV}^{-2}}$ and $\hat{\alpha}_s = 0.1185$.
The expressions for the $x^{(1)}(\mu)$ used are summarized in Ref.~\cite{Buttazzo:2013uya}. The value indicated for the QCD
coupling $g_3$ includes RG running at 4-loops in QCD and 2-loops in the EW
interactions employing the expression as a function of $\alpha_3(m_Z)$ and $m_t$ as in
Ref.~\cite{Buttazzo:2013uya}.
The quantity $\lambda(\mu=m_t)$ does not show significant dependence on the parameter $|\omega|$:
for any value $|\omega| < 0.1$ we have $|\Delta\lambda| \lesssim 10^{-6}$, which is numerically insignificant
in the running. Conversely, this quantity is quite sensitive to the scale $m_p$ and the RGE order used.
There is significant numerical sensitivity to the input parameters.
In particular, the precise experimental determination of $\{\hat{m_h},\hat{m_t}\}$ is critical
for the consistency of the scenario. We show the dependence on these inputs
on the inferred scale $m_p$ in Fig.~\ref{fig:lambda_vs_M}. This plot shows the value for $\lambda(\hat{m}_t)$ consistent with experimental measurements is obtained for $m_p\simeq 10^{1.3}$~PeV, assuming $\hat{m}_t=173.2$~GeV.
\begin{figure}
 \includegraphics[width=0.5\textwidth]{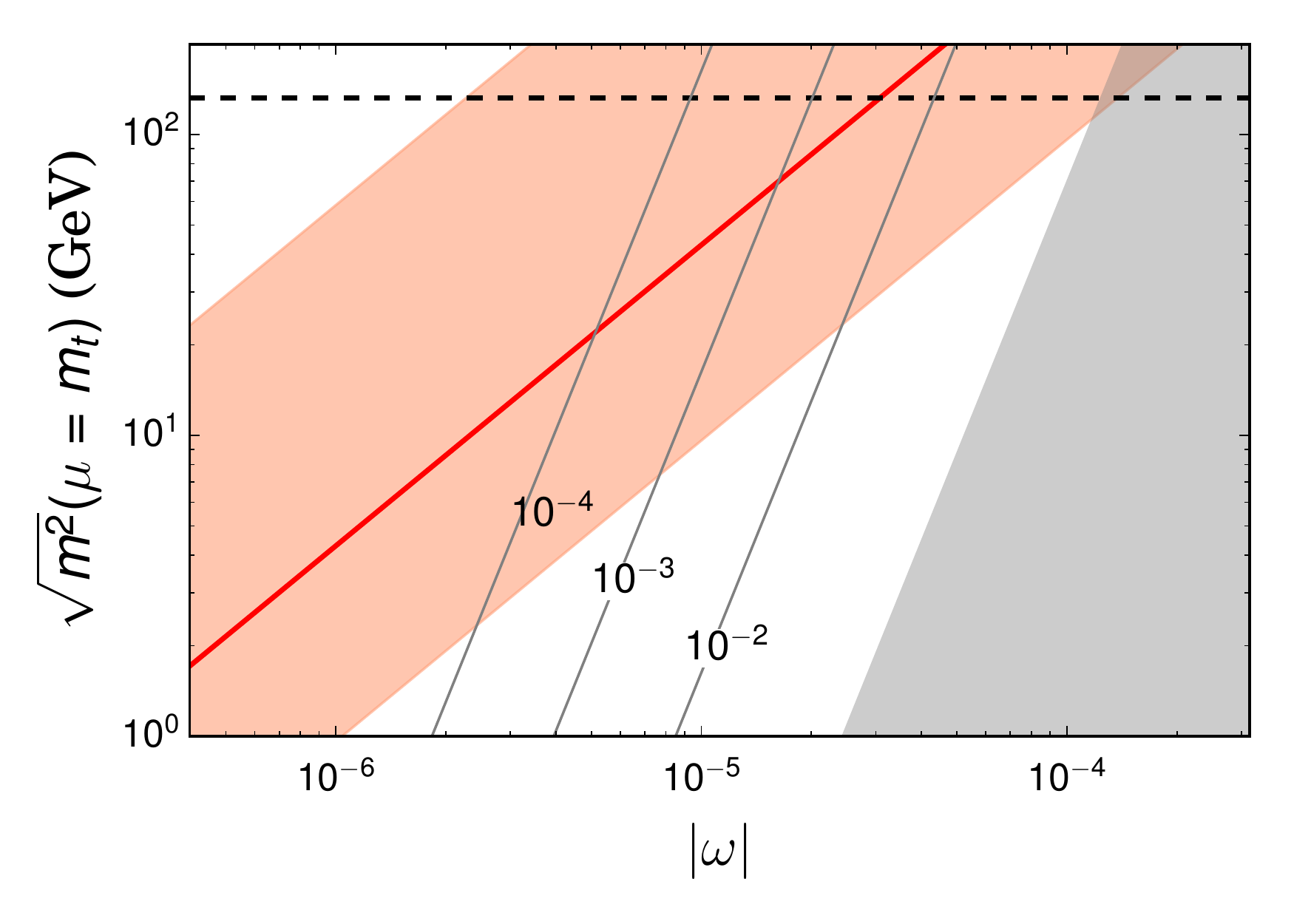}
 \caption{\label{fig:omega_vs_M}
 Value of $\sqrt{m^2}$ extrapolated at $\mu=\hat{m}_t$ as a function of $|\omega|$. The dashed black horizontal line indicates the value consistent with the measured Higgs mass,
 while the red solid line is obtained for $m_p = 10^{1.3}$ PeV. 
 The red shaded region corresponds to the uncertainty on the top quark mass, consistent with Fig.~\ref{fig:lambda_vs_M}
 and the grey region is disfavoured due to $\Lambda$ CDM cosmology limits on the sum of neutrino masses (Eqn.~\ref{limits}). The neutrino mass scales predicted (in ${\rm eV}$) are the three solid lines.}
\end{figure}
The quantity $m^2(\mu=\hat{m}_t)$ is sensitive to both $m_p$ and $|\omega|$. Fig.~\ref{fig:omega_vs_M}
shows its dependence on $|\omega|$ for the fixed value $m_p=10^{1.3}$~PeV
 and the corresponding viable band associated to the uncertainty on the top
 mass determination.\footnote{This interesting region of parameter space in the seesaw model has been
 previously discussed in Ref.~\cite{Vissani:1997ys}.}

\begin{figure}
\includegraphics[width=0.5\textwidth]{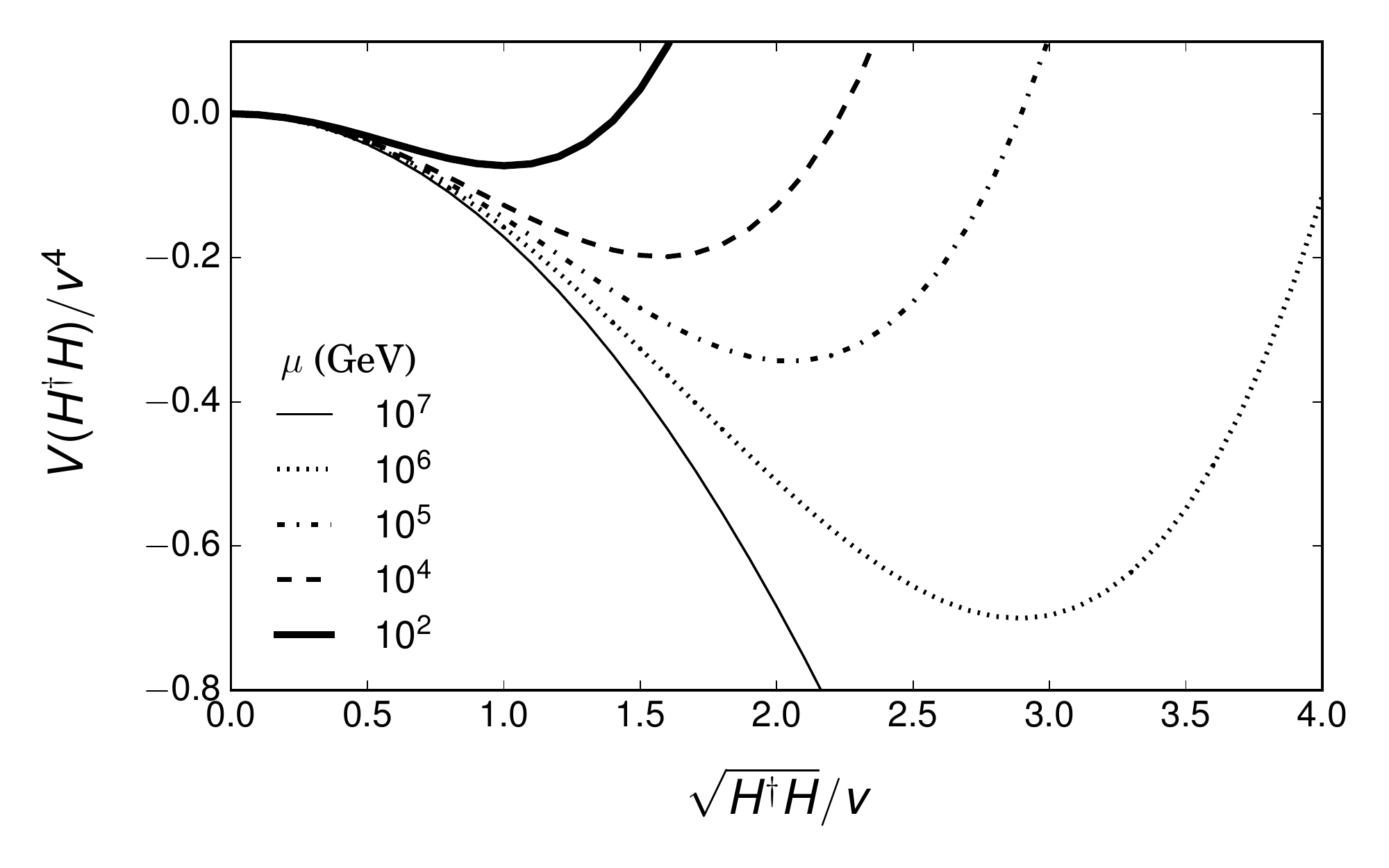}
\caption{\label{fig:wowplot}  The emergence of the Higgs potential due to
running the seesaw boundary conditions down to $\mu \sim \hat{m}_t$.}
\end{figure}

This scenario can lead to a
SM-like Higgs potential emerging from the combined effect of the threshold corrections and
the SM RGEs in a non-trivial fashion as shown in Fig.~\ref{fig:wowplot}.

\paragraph{\bf V. Cosmological and low energy constraints.}
The sum of the observed neutrino masses is  $\sum_i m_\nu^i \simeq 3 \, |\omega|^2/2 \, \sqrt{2} \hat{G}_F \, m_p$
in the tree level approximation used here, while neglecting running effects. Assuming a $\Lambda$ CDM cosmology,
combined CMB, supernovae and Baryon Acoustic Oscillation data limits this sum \cite{Ade:2015xua}.
This translates into a constraint of
\bea\label{limits}
\frac{3 \, \sqrt{3}}{8 \, \pi} \, \frac{ |\omega|^2}{\hat{G}_F \, \sqrt{\Delta m^2}} \lesssim 0.23 \, {\rm eV},  \quad \quad {\rm 95 \% C.L.},
\eea
which is shown as a grey exclusion region on Fig.~\ref{fig:omega_vs_M}.

The overall neutrino mass scale predicted is very sensitive
to the uncertainty on $\hat{m}_t$, the chosen order of RGEs,
and threshold loop corrections included in the numerical simulation.
In Fig.~\ref{fig:omega_vs_M}
we show the absolute
neutrino mass scales (grey lines) predicted at leading order as $|m_\nu| = 3 \, |\omega|^2 /2 \, \sqrt{2} \, \hat{G}_F \, m_p$.
One expects
$|m_\nu|^2 \gtrsim \Delta m^2_{21},\Delta m^2$ to avoid fine tuning and a requirement of further model building
in the Majorana sector.

In addition, a negative sign for $\lambda$ and $m^2$ indicates a theory with a Hamiltonian unbounded from below. However, the corresponding decay time for the EW vacuum is exponentially small
\cite{Kobzarev:1974cp,Coleman:1977py,Callan:1977pt,Isidori:2001bm}.
We have checked that the EW vacuum decays in this scenario are well
approximated by the (negligible) result in the SM in Ref.~\cite{Buttazzo:2013uya}. The ratio of the scales at which
$\beta(\lambda)$ vanishes in the SM, compared to the SM extension considered in this paper,
(which fixes the size of the action of the bounce) is $\sim$ 1.00011.
The extrapolation of the theory far above the scale $m_p$ is associated with a large theory uncertainty
as the $N_p$ could be embedded in an extended Majorana sector,
with other states that can also modify the running of the couplings above the scale $\mu \simeq m_p$.

\begin{figure}
\includegraphics[width=0.45\textwidth]{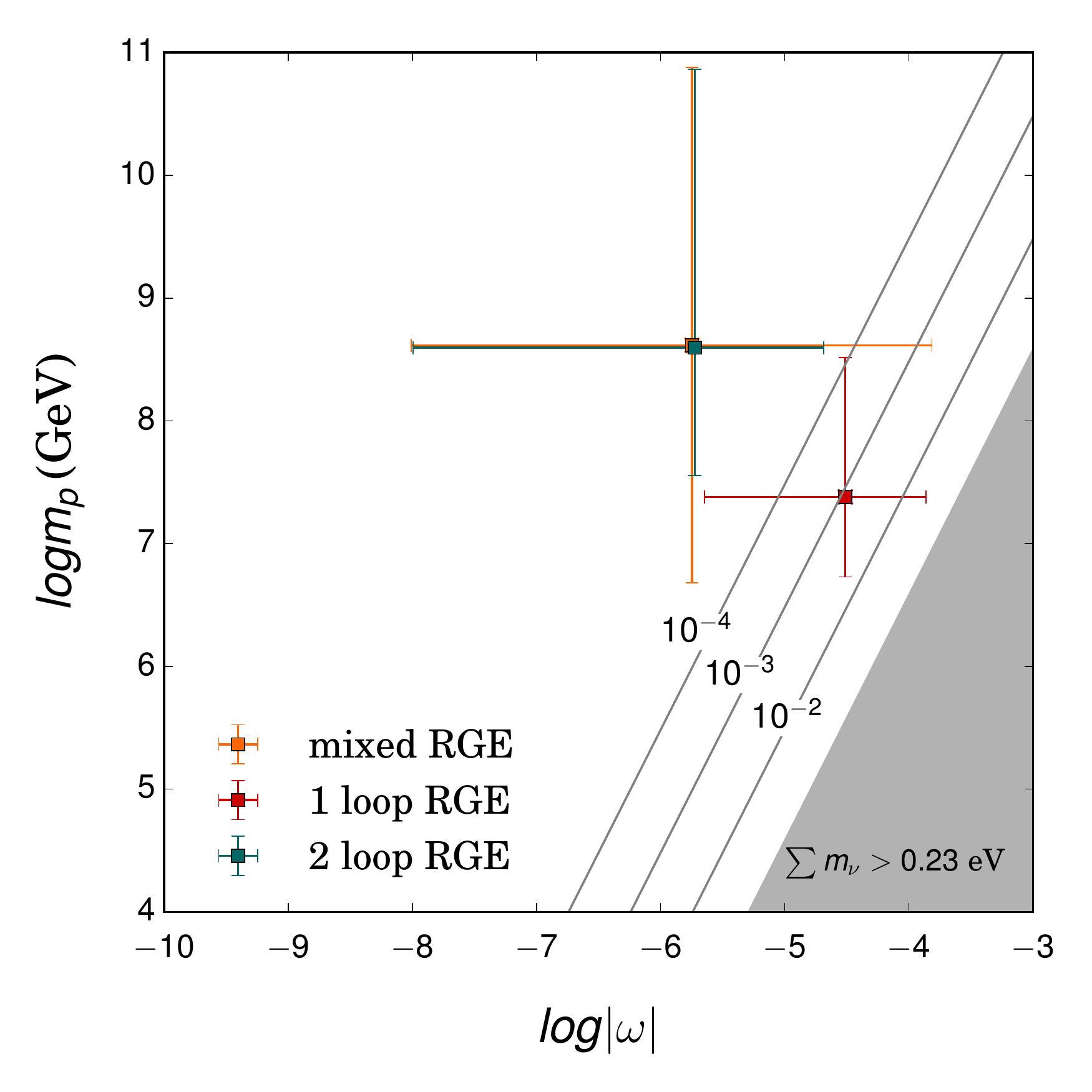}
\caption{\label{unburied} Numerical sensitivity of the results, with all cases
showing one loop matching to $\Delta m$, $\lambda$ (due to the model assumption)
and including one loop corrections for the remaining SM parameters with
one, two loop, or mixed RGEs for all parameters. The mixed case shows one loop running
for $\Delta m$, $\lambda$ and two loop running for all remaining SM parameters. The best fit points
are indicated with a box in each case with error bars showing the experimental uncertainty in the top quark
mass, which has been chosen to be its $2 \, \sigma$ uncertainty \cite{Olive:2016xmw}. Nevertheless, we have determined that
the measured mass differences of the neutrinos can be accomodated in all three RGE cases.}
\end{figure}

\paragraph{\bf VI. Numerical stability of the results.}
The results shown in Section IV are produced with one loop matching conditions
and one loop RGEs. Increasing the RGE and threshold matching order used shows significant numerical sensitivity.
This is essentially because the coupling $\lambda$ is running to small and negative values
asymptotically which introduces a sensitivity to the scale $m_p$ where the
seesaw boundary conditions are matched. This feeds into the required $|\omega|$ to produce the
Higgs potential and EW scale, and subsequently the neutrino mass scale.
For this reason, the minimal scenario is falsifiable.
On the other hand, the uncertainty in $\hat{m}_t$ is significant.
To illustrate this we show in Fig.~\ref{unburied} the best fit points for the cases where the boundary conditions of the
scenario are evolved with one loop SM RGEs, two loop SM RGEs, and one loop RGEs for
$\Delta m$ and $\lambda$ and two loop RGEs for the remaining SM parameters.\footnote{Formally the running should be described using
the SMEFT RGEs which include the effect of higher dimensional operators feeding into the running of the SM couplings \cite{Jenkins:2013zja}.
We have checked that this effect is numerically sub-dominant in this model and neglected it.} This last case is
shown as these parameters do not have a tree level matching coefficient in
this scenario.
Despite this, we have confirmed that using one loop or two loop RGE's the measured neutrino mass
differences can be reproduced, see Section VIII.

\paragraph{\bf VII. Tree level decays and IceCube.}
The tree level decays of the $N_p$ states are well known, see Refs.~\cite{Dolgov:2000jw,Gorbunov:2007ak}.
Remarkably, the mass range selected for when the Higgs potential is radiatively generated in the minimal seesaw scenario is consistent with the measured
energies of an excess of neutrinos reported by IceCube \cite{Aartsen:2013bka,Aartsen:2013jdh,Aartsen:2014gkd}.
The $d \Gamma/ d E_\nu$ spectrum that results from these decays
is a sharp mono-chromatic peak at the scale $m_p/2$.

The possibility that the $N_p$ states can be viable dark matter candidates to induce the IceCube
events has been examined in the literature. We agree with Refs.~\cite{Feldstein:2013kka,Esmaili:2013gha,Esmaili:2014rma}
that the required coupling for the event rate scales as $\Gamma_{events} \sim (|\omega|/10^{-29})^2 (m_p/1.2 {\rm PeV})/year$
which is inconsistent  with the preferred $|\omega|$ to generate the Higgs potential in the minimal seesaw model with a
PeV Majorana sector. Extended model building can possibly accommodate these observations.

\paragraph{\bf VIII. Neutrino mass differences and mixing.}\label{massesmixing}
The measured mass differences for the neutrino masses are given as \cite{Olive:2016xmw}
\bea
\Delta m^2_{21}/10^{-5} {\rm eV^2} &=&  6.93 - 7.97, \\
\Delta m^2/10^{-3} {\rm eV^2} &=&  2.37 - 2.63 \, (2.33 - 2.60)
\eea
using PDG notation. Here the quoted range is $3 \, \sigma$ and the brackets indicate an inverted
$m_\nu$ hierarchy. In our simple approximation of a common universal scale and coupling
for the $N_p$ states integrated out, we do not predict mass differences of the
$m_\nu$ or mixing angles. 
Treating $N_p$, $\omega$ and the charged lepton Yukawa $Y_e$ as non-degenerate enlarges the number of free parameters
(15 moduli + 6 phases~\cite{Broncano:2002rw}) to a set larger than that of the experimental constraints (2 Higgs parameters + 2 $\nu$ mass differences + 3 PMNS angles).
As an existence proof, consider $\omega \equiv \omega_0 \mathbbm{1}+\delta \omega$ with $(\delta\omega)_{ij}\ll \omega_0$. Correct values for $m^2(\hat m_t)$, $\lambda(\hat m_t)$, $\Delta m_{12}^2$ and $\Delta m^2$ (see Ref.~\cite{Olive:2016xmw})
in a normal hierarchy can be obtained e.g. for $\{m_1, m_2, m_3\} = \{23.96, 24.77,25.27 \}$~PeV, $\omega_0=10^{-4.4}$, $\delta\omega_{11}= 4.08 \cdot 10^{-10}$,
$\delta\omega_{22}=-1.88 \cdot 10^{-8}$ and $\delta\omega_{33}=-7.67 \cdot 10^{-9}$. Only the diagonal terms in $\omega$ were used in this example leading to $U_\nu = {\rm diag}\{1,1,1\}$.
The results can be perturbed by the free off diagonal entries that can combine with the unfixed charged lepton mass matrix rotation matrix $U_\ell$ so that
$U_\ell^\dagger \, U_\nu = U_{PMNS}$.

\paragraph{\bf IX. Conclusions.}
Due to a non-trivial interplay of the couplings of the SM and the mass scales of the SM states expanded around the classically scaleless limit,
the minimal seesaw scenario can form a UV boundary condition that induces the Higgs potential at lower energies.
This can occur as a simple mechanism to generate neutrino masses is introduced extending the SM.

In this scenario the EW scale is not fundamental but is due to the quantum threshold corrections matching the heavy singlet states onto the SMEFT, which is assumed to be expanded in its near scaleless limit.
Instead of an expectation of new states at the TeV scale, the multi-PeV scale is the locus of a requirement of a mass generating mechanism
for Majorana states, and possibly accompanying stabilizing symmetries such as supersymmetry.
This change in perspective on the hierarchy problem is possibly valuable. The key point underlying
this approach is to abandon attempts to stabilise the Higgs mass against threshold corrections at the TeV scale, due to the lack of experimental indications
of new states associated with stabilizing symmetries. Instead we advocate embracing these corrections
as the origin of the Higgs potential. This approach can also be developed in other models beyond the minimal scenario considered here.

Future experimental results supporting this scenario are a continued lack of discovery of states motivated by a traditional interpretation of
the hierarchy problem at the LHC,
and the eventual discovery of the Majorana nature of neutrinos in $0\nu \beta \beta$ decay. The scenario can also be tested through consistency tests
of the neutrino mass spectrum and the PMNS matrix due to a minimal seesaw scenario and more precise measurements of $\hat{m}_t,\hat{m}_h$ at LHC. In this manner the Neutrino Option in generating the Higgs potential is falsifiable.
Further phenomenological investigations, and the advance of higher order
SM threshold and RGE calculations are also strongly motivated.

{\bf Acknowledgements}
MT and IB acknowledge generous support from the Villum Fonden and partial support by the Danish National Research Foundation (DNRF91).
We thank Cliff Burgess, Jacob Bourjaily, Poul Damgaard, Gitte Elgaard-Clausen, Bel\'en Gavela, Yun Jiang, Jason Koskinen, David McGady,
Philipp Mertsch, Andr\'e Mendes and  Jure Zupan for helpful discussions,
and Andr\'e David for technical assistance. MT thanks the organizers of EW-Moriond 2017 for an inspiring environment as
this paper was completed.
\vspace{-0.7cm}

\end{document}